\documentclass[prd,nofootinbib,floats,superscriptaddress,eqsecnum,tightenlines,11pt]{revtex4}

\usepackage{hyperref}
\usepackage{graphicx}
\usepackage{amsmath,amssymb,amsfonts,amsthm,latexsym,stmaryrd}
\usepackage{marginnote}
\usepackage{color}

\def\be{\begin{equation}}
\def\ee{\end{equation}}
\def\ba{\begin{eqnarray}}
\def\ea{\end{eqnarray}}
\def\bas{\begin{subequations}\begin{eqnarray}}
\def\eas{\end{eqnarray}\end{subequations}}

\begin{document}

\title{Analytic continuation of real Loop Quantum Gravity : \\
Lessons from black hole thermodynamics}

\author{Jibril Ben Achour }
\email{benachou@apc.univ-paris7.fr}
\affiliation{Laboratoire APC -- Astroparticule et Cosmologie, Universit\'e Paris Diderot Paris 7, 75013 Paris, France}

\author{Karim Noui}
\email{karim.noui@lmpt.univ-tours.fr}
\affiliation{Laboratoire de Math\'ematiques et Physique Th\'eorique, Universit\'e Fran\c cois Rabelais, Parc de Grandmont, 37200 Tours, France}
\affiliation{Laboratoire APC -- Astroparticule et Cosmologie, Universit\'e Paris Diderot Paris 7, 75013 Paris, France}

\begin{abstract}
This contribution is devoted to summarize the recent results obtained in the construction of an ``analytic continuation" of Loop Quantum Gravity (LQG). By this, we mean that we construct analytic continuation
of  physical quantities in LQG  from real values of the Barbero-Immirzi parameter $\gamma$  to the purely imaginary value $\gamma = \pm i$. This should allow us to define a quantization of gravity with self-dual Ashtekar variables. 
We first realized  in \cite{Geiller1} 
that this procedure, when applied to compute the entropy of a spherical black hole in LQG for $\gamma=\pm i$,  allows to reproduce exactly the Bekenstein-Hawking area law at the semi-classical limit.
The rigorous construction of the analytic continuation of spherical black hole entropy has been done in  \cite{Achour1}. 
Here, we start with a review of  the main steps of this construction: we recall that our prescription turns out to be unique  (under natural assumptions) and  leads to the right semi-classical limit with its logarithmic quantum corrections. Furthermore, 
the discrete and $\gamma$-dependent area spectrum of the black hole horizon becomes continuous and obviously $\gamma$-independent. Then, we review how this analytic continuation could be interpreted in terms of an analytic continuation
from the compact gauge group $SU(2)$ to the non-compact gauge group $SU(1,1)$ relying on an analysis of three dimensional quantum gravity.
\end{abstract}

		

\maketitle

\section{Motivation: getting rid of $\gamma$}
The Barbero-Immirzi parameter $\gamma$ seems to play  apparently a paradoxical role in LQG. Whereas it is totally irrelevant in the classical theory, it enters into the expressions of  ``physical"  quantities 
like eigenvalues of geometric operators in the kinematical sector, the maximal density of the Universe in quantum cosmology or the black hole entropy. Nonetheless, many observations (from black holes physics \cite{Geiller1, Achour1} and three dimensional
quantum gravity \cite{Achour3, Achour4})  indicate that $\gamma$ should somehow ``disappear" from the quantum theory in the sense that it should take the natural complex value $\gamma=\pm i$ and not a real value. We review here some of
these observations  focussing mainly on the case of the black hole entropy. We finish with a discussion where we quickly review  the state of  the art in three dimensions. 

\section{Analytic continuation of black hole entropy}
\subsection{Real $\gamma$ Black Holes}
In the framework of LQG, a black hole is defined as a boundary in space-time which satisfies the constraints of an isolated horizon \cite{Abay}.
Those constraints impose that the black hole degrees of freedom are encoded into the phase space of an $SU(2)$ Chern Simons theory defined  on a punctured two-sphere $S^{2}$ as a canonical surface.
The presence of the gauge group $SU(2)$ derived from the phase space of gravity in the bulk expressed in terms of Ashtekar-Barbero variables. 
The Chern-Simons level $k$ is proportional to the horizon area $a_H$ and depends on $\gamma$ according to (\ref{area and level})\footnote{Variant expressions exist but the precise dependence on $\gamma$ is not important
for our purpose. The main point is that $k$ is large when the area $a_H$ is macroscopic.}.
The punctures originate from the spin networks (the quantum states of the gravitional field) defined in the bulk which pierce the horizon. They are viewed as the fundamental excitations of the black hole and each puncture 
carries a quantum of area $a_{l}$ which contributes to the macroscopic area  $a_{H} $ in the usual ``real $\gamma$" picture according to ($n$ labels the number of punctures)
\begin{equation}\label{area and level}
a_{H} = \frac{2\pi \gamma }{(1-\gamma^{2})}k  = \sum^{n}_{l=1} a_{l} \;\;\;\;\;\;\;\;\text{with}\;\;\;\;\;\;\;\;   a_{l} = 8 \pi l^{2}_{P} \gamma  \sqrt{j_{l}(j_{l}+1)} . 
\end{equation}
As usual in LQG, the spin $j_{l} \in \mathbb{N}/2$ labels an $SU(2)$ unitary irreducible representation (irrep).
For a fixed $n$, a microscopic state of the black hole is  defined by an ordered\footnote{In the usual real picture, the punctures are distinguishable. For this reason, we consider a priori
an ordered family of spins. In the complex picture, we relax the distinguishability.} family of spins  $\mathcal{P} = (j_{1}, ...., j_{n})$. 

The degeneracy of a configuration $\mathcal{P}$  is given by the dimension of
the Chern-Simons Hilbert space  $\mathcal{H}_{k}(S^2;j_1,\cdots,j_n)$ which is well-known to be defined by the space of $U_{q}(su(2))$ invariant tensors\footnote{The space of invariant tensors is
endowed with the quantum Haar measure when viewed as the space of linear forms on $SU_q(2)$, the polynomials of the quantum deformation of $SU(2)$.}
in the tensor product $\otimes_l V_l$. Here $V_l$ are  $U_q(su(2))$ modules labelled by the spins $j_l$ whose dimension is denoted $d_l=2j_l+1$.
The quantum parameter is a root of unity defined by $q=\exp(i\pi/(k+2))$ where the level $k$ is necessarily integer. The dimension $g_{k}(d_{l})$ of this Hilbert space is easily computed \cite{Karim1} and can be expressed as
the  following sum over the integer $d$
\begin{equation}\label{Verlinde}
g_{k}(d_{l}) = \frac{2}{2+k} \sum^{k+1}_{d=1} \sin^{2}(\frac{\pi d}{k+2}) \prod^{n}_{l=1} \frac{\sin(\frac{\pi}{k+2}dd_{l})}{\sin(\frac{\pi }{k+2}d)}.
\end{equation}
This Verlinde formula allows to recover at the semi-classical limit ($a_H$ large in Planck units) the Bekenstein-Hawking  law for the black hole entropy provided that $\gamma$ is fixed to a peculiar value \cite{Rov1, Ash1}. 
This is easily seen in the simplest model  where $d_l=d$ for any $l$ is fixed and   $n$  becomes large at the semi-classical limit.  Even if this result is certainly an important  success of LQG, it has risen 
important questions concerning the role of $\gamma$ and the validity of the computation we have just sketched.  Since then, different interpretations of $\gamma$ have been discussed but none are totally
convincing (see \cite{GeillerNoui} and references therein).

\subsection{Complex $\gamma=\pm i$ Black Holes}
The last couple of years, a new road towards the understanding of the role of $\gamma$ has emerged. 
In that new picture, the Barbero-Immirzi parameter is viewed as a ``regulator" which should be sent back to its ``original'' imaginary value $\gamma=\pm i$. To clarify this point of view, let us recall that $\gamma$ has been first introduced to overcome the problem of working with complex variables, to circumvent the resolution of the reality conditions and then to start the loop quantization of gravity. It is important to notice that such a strategy has been successful because it has led to a very beautiful
picture of the quantum (kinematical) geometry  at the Planck scale. However, for solving the quantum dynamics, the real Ashtekar-Barbero connection doesn't seem to be well suited anymore. Already at the classical level, it is well known that this connection 
doesn't transform properly under timelike diffeomorphisms, and this might be the reason why $\gamma$ remains in the theory at least at the kinematical level. 
This fact is enhanced by a series of recent works which all point towards the need to come back to the (anti) self-dual variables \cite{Geiller1,Dan1,Dan2,Dan3, GeillerNoui2}.  
One of the most striking result \cite{Geiller1,Achour1} in that respect has been obtained in the context of black hole physics: the analytic continuation of the formula (\ref{Verlinde}) for the Chern-Simons Hilbert space dimension
to the imaginary value $\gamma = \pm i$ allows to reproduce the expected semi-classical Bekenstein-Hawking law for the black hole entropy. We are  going to briefly recall how this works following the construction of \cite{Achour1}. 
Details can be found in the original paper \cite{Achour1}.

First of all, we immediately notice that taking $\gamma = \pm i$  leads to a complex value of the Chern-Simons level which becomes $k=i\lambda$ with $\lambda \in \mathbb R$. From the LQG point of view,
this is an immediate consequence of (\ref{area and level}).
From the Chern-Simons theory point of view, this shift from $k \in \mathbb{N}$ to $k \in i \mathbb{R}$ can be interpreted by the fact that one works now with a complex $SL(2,\mathbb C)$ connection rather than a compact real 
$SU(2)$ connection \cite{Witten1}. Unfortunately, Chern-Simons theory with complex gauge group and complex level is poorly understood at the quantum level and the only  one serious candidates for its quantization deeply relies on analytic 
continuation techniques \cite{Witten1}.  However, the process of analytic continuation is rather subtle even in the construction of the analytic continuation of the Hilbert space dimension (\ref{Verlinde}).  
Indeed, since $k$ enters in the upper bound of the sum, the expression (\ref{Verlinde}) is not really convenient for analytic continuation purposes even if we used it formally in the first proposal \cite{Geiller1}.  It is much more  
convenient to view (\ref{Verlinde}) as a sum of  residues of an analytic function in order to write it  as an integral in the complex plane along a contour $\mathcal{C}$ (see figure \ref{contour}) which encompass the imaginary axis between $[0, i\pi]$:
\begin{figure}
\includegraphics{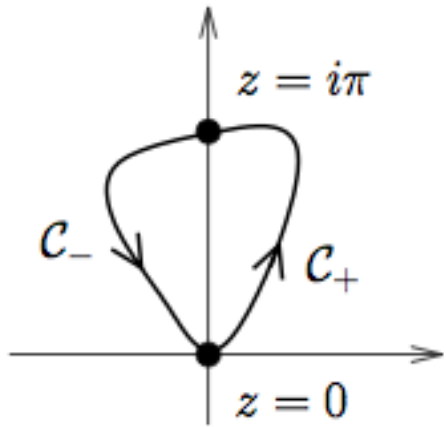}
\caption{The contour $\mathcal{C} = \mathcal{C}_{+}+ \mathcal{C}_{-} $ in the complex plane encloses the imaginary axis between $[0, i\; \pi]$.}
\label{contour}
\end{figure}
\begin{equation}\label{complex integral}
g_{k}(d_{l}) = \frac{i}{\pi} \oint_{\mathcal{C}} \; dz \; \sinh^{2}(z) \; \prod^{n}_{l=1} \frac{\sinh(d_{l} z)}{\sinh(z)} \; \coth((k+2)z).
\end{equation}
To simplify the discussion below, we introduce the notation $G(z) =  \coth((k+2)z)$ while the remaining part of the integrand will be denoted $F(z)$.
When $k$ and $d_{l}$ are integers, the poles of the integrand are the poles of $G(z)$ which are located on the imaginary axis: $z_{p} = \frac{i \pi p}{k+2}$ for $p \in \mathbb{N}^{*}$. This justifies the choice of the contour $\mathcal{C}$.

\medskip

Now, it makes sense to consider  $k = i \lambda$ as $G$ is an analytic function of $k$. Let us discuss what happens to the integral (\ref{complex integral})  when one performs such an analytic continuation. 
\begin{itemize}
\item If the dimensions $d_{l}$ remain integer, the poles of the integrand are located on the real axis: 
the poles of  $G(z)$ are  $z_{p} =  - \frac{ \pi p}{\lambda}$ with $p \in \mathbb{N}^{*}$ whereas $F(z)$ still has no pole. In this case, a contour $\mathcal{C}$ close enough to the complex axis doesn't enclose any poles and the integral vanishes.
As a consequence,  analytic continuing only the Chern Simons level from $k \in \mathbb{N}^{*}$ to $k \in i \mathbb{R}$ keeping $d_l$ unchanged leads to inconsistent physical results.
\item If $d_{l} = i s_{l}\in i\mathbb R$, the analytic continuation is much more interesting. 
The location of the poles of $G(z)$ is unchanged compared to the previous case but the novelty is that $F(z)$ admits new poles on the imaginary axis at $z_{m} = i \pi m$ with $m \in \mathbb{N}^{*}$. 
Among all these new poles, we are more interested in the one located at $i\pi$ which must be enclosed by the contour $\mathcal{C}$ in order for the integral (\ref{complex integral}) to be non-trivial.
Therefore, such a continuation leads to a non-trivial result for the black hole Hilbert space that we define to be the dimension\footnote{To be interpreted as a dimension, $g_k$ must necessary be  a non-negative real number. This is asymptotically
the case, i.e. when the horizon area becomes large in Planck units and under some conditions satisfied by the number $n$ of punctures. 
For non large area, $g_k$ is in general complex but we can argue that we have to consider $\vert g_k \vert$ or ${\mathfrak R}(g_k)$  as the dimension
even if this aspect deserves to be studied deeper. In \cite{Achour1}, we considered the modulus of $g_k$.}  of the black hole Hilbert space when $\gamma = \pm i$.
\end{itemize}
To summarize, the dimension  of the black hole Hilbert space for $\gamma = \pm i$ is defined from the analytic continuation of (\ref{complex integral}) when $\mathcal{C}$ encloses the point $i\pi$
and
\begin{equation}
 k= i\lambda \;\;\;\;\; ,   \;\;\;\;\; d_{l} = i s_{l}  \Leftrightarrow j_{l} = \frac{1}{2} (-1 + i s_{l}) \;\;\;\text{with}\;\;\;\;  \lambda,s_{l} \in \mathbb{R}^{+} \,.
\end{equation}
The fact that $\lambda$ and $s_l$ are non negative is not restrictive.  At this point,  it is important to explain the choice $d_l \in i  \mathbb R$. From a physical point of view,
this is the only consistent choice which leaves the area spectrum real when $\gamma=\pm i$ as seen from 
\begin{equation}
A(j_l) = 8\pi l^{2}_{p} \gamma  \sqrt{j_l(j_l+1)}  \;\;\;\;    \stackrel{\gamma=\pm i}{{\longrightarrow}} \;\;\;\;\;  A(s_l) = 4 \pi l^{2}_{p} \sqrt{s_l^{2} + 1}\
\end{equation}
where we choose the square root of $-1$ such that the area is non negative.
From a  mathematical point of view, changing $j_l$ to $\frac{1}{2} (-1 + i s_{l})$ amounts to considering $SU(1,1)$ irreps instead of $SU(2)$ irreps for coloring the punctures.
 
 \subsection{Semi-classical limit: area law and logarithmic corrections}
To simplify the study of the semi-classical limit, we consider the model where all the punctures carry the same color $s_l=s$. This corresponds to the one color model in the following. 
We first impose that $k$ is large in (\ref{complex integral}) and we obtain the following expression for the candidate to  the dimension of the one-color black  hole Hilbert space
\begin{equation}\label{one color}
g_{\infty}(s, n) = \frac{i}{\pi} \oint_{\mathrm{C}} \; dz \; \sinh^{2}(z) \; e^{n S(z)}  \;\;\;\;\;\text{with}   \;\;\;\;\;  S(z) =  \log \big{(}\frac{\sinh(s z)}{\sinh(z)} \big{)}.
\end{equation}
In the semi-classical limit, the black hole area $a_H = 4\pi \l^2_p ns$ is large, which means that the product $ns$ becomes large. 
It has been argued in \cite{Gosh3} that the semi-classical regime corresponds to both $n$ and $s$ large.
Then, the form (\ref{one color}) of the integral is well suited for the study of the thermodynamical limit. 
When $n$ is large, this integral can be estimated using the stationary phase method. The study of the critical points reveals that there are two critical points, $z_{c}=0$ and $z_{c} = i(\pi + \frac{1}{s}) + {o}(\frac{1}{s})$ for $s$ large. 
Only the later contributes to the saddle point approximation which finally leads to
\begin{equation}
S_{m} = \log(g_{\infty}(s, n) ) =  \frac{a_{H}}{4l^{2}_{p}} + S_{cor}   
\end{equation}
for the black hole microcanonical entropy $S_m$.
The leading term reproduces the expected Bekenstein Hawking area law  without any fine tuning and $S_{cor}$
are quantum corrections. At this point, the quantum corrections scale in general as $\sqrt{a_H}$ in Planck units, and then are much larger than logarithmic corrections.

In the grand canonical ensemble, the situation concerning the quantum corrections is more satisfying.
Using the local framework developed in \cite{Gosh1,Gosh2}, we have a notion of energy (measured by an observer located at a ``small" distance $L$ of the horizon compared to the black role radius) 
which allows to compute the canonical and grand canonical partition functions.
In this approach, the black hole is viewed as a gas of indistinguishable punctures which has been studied first in \cite{Gosh3}. If we assume in addition  that the punctures admit a non-vanishing chemical potential $\mu$
and satisfy the Maxwell-Botzmann statistics, we can show that, at the semi-classical limit, the black hole temperature approaches the Unruh temperature $T_U=1/\beta_U$ for the local observer and the black hole mean area $\bar{a}_H$,
the mean number of punctures $\bar{n}$ and the mean color $\bar{s}$ scale as follows
\begin{equation}
\bar{a}_{H} = 4 \pi l^{2}_{p}\frac{z}{2x^{2}}  \left(1- \frac{3}{2z}x\right)  \;\;\;\text{with}\;\;\;  x = \frac{2L}{l^{2}_{p}} ( \beta - \beta_{U})  \;\;\;\;\text{and}\;\;\;  \bar{n} \propto \sqrt{a_{H}} \;\;,\;\;\;  \bar{s} \propto \sqrt{a_{H}} \;\;\;\;\;\;\;\;   
\end{equation}
where $z=\exp(\beta\mu)$ is the fugacity. As a consistency check, we recover that $n$ and $s$ are large in the semi-classical regime.
It is easy to compute  from this analysis the semi-classical expansion of the grand canonical entropy and we deduce the expression
\begin{equation}\label{gc entropy}
S_{gc} =  \frac{\bar{a}_{H}}{4l^{2}_{p}} - \frac{3}{2}\log \left( \frac{\bar{a}_{H}}{l^{2}_{p}}\right)  + \frac{z_{U}}{2}(2-\mu\beta_{U}) \left(\frac{\bar{a}_{H}}{l^2_p}\right)^{1/2} + o\left( \log \left( \frac{\bar{a}_{H}}{l^{2}_{p}}\right)\right).   
\end{equation}
We recover therefore the expected logarithmic quantum corrections supplemented with larger quantum corrections $\propto \sqrt{a_{H}}$ which vanish when the chemical potential is fixed to $\mu = 2T_{U}$.  
A physical interpretation of such a result is still missing. Note however that the same value of the chemical potential is also found to cancel the too large quantum correction for the real black hole ($\gamma \in \mathbb{R}$) with the Maxwell Boltzman statistic \cite{Achour2}. Therefore, this behaviour of the quantum correction $S_{cor}\propto \sqrt{a_{H}}$ is not specific to the complex model.

The calculation can be generalized in different ways.  First, we can extend the model to a black hole with $p$ colors \cite{Achour1}. In that case, the entropy has the same form as (\ref{gc entropy}) 
with a modified logarithmic correction which depends on $p$. Interestingly we notice  that only the case $p=1$, i.e. the one color model, which can be interpreted as a kind of spherical symmetric quantum condition,
 allows to recover the prefactor $-3/2$ for the logarithmic corrections. We could also generalize to the cases where the punctures satisfy a quantum statistic. This has been done in \cite{Gosh3,Achour2} when the area spectrum is
 discrete.  In that context, we showed that assuming  the punctures are boson, there exists a semi-classical regime where the gas condensates to spin $1/2$ punctures and also where the quantum corrections
 are logarithmic\footnote{This contrasts with the case of a classical Maxwell-Boltzmann statistics where large spins dominate at the semi-classical limit.}. 
  It would be interesting to see whether a similar phenomenon occurs when the spectrum is continuous.

\section{Discussion: 2+1 dimensional gravity as a guide} 
We have quickly reviewed the analytic continuation procedure  applied to black hole entropy first proposed in \cite{Geiller1} and then more  rigorously developed in \cite{Achour1}.
This procedure turns out to give an unique candidate for the dimension of the complex $\gamma=\pm i$ black hole Hilbert space. Asymptotically, this candidate allows to reproduce 
the Bekenstein-Hawking law for the black hole entropy. 

\subsection{Open issues}
However, even if the result is really striking, many aspects remains to be clarified. First, the complex dimension is only ``asymptotically" positive and is in general complex for non-large area.  Then we have
to argue that one should consider its modulus or its (positive) real part as a good definition of the complex dimension.  Furthermore, our analysis is somehow only an observation and it
is now important to understand  the deep meaning of this analytic continuation. We are currently working in that direction.
 It seems that the gauge group $SU(1,1)$ plays a crucial role and replaces $SU(2)$ when we get rid of the Barbero-Immirzi parameter.
Indeed, we recognize that the new area spectrum (at least for a black hole horizon) is given by the Casimir of $SU(1,1)$  in the continuous series. Should $SU(1,1)$  play also an important role in the full theory?
If this is the case, how the kinematical $SU(2)$-spin networks are modified once one applies the analytic continuation prescription ? Those questions remain largely open and really important to study. 

\subsection{Guided by three-dimensional gravity}
Nonetheless, we can be guided by an interesting result  derived in the context of three dimensional gravity \cite{Achour3,Achour4}. Indeed, we have shown that it is possible to construct a three dimensional analog of the Holst action
which consists in an action for (Lorentzian or Euclidean) three-dimensional gravity supplemented with a $\gamma$-dependent term exactly as in four dimensions. We will concentrate only on the Lorentzian
action in this brief review.  In fact, such an action can be obtained from a symmetry reduction
of the four-dimensional Holst action to three dimensions and can be written explicitly as follows:
\begin{equation}
S_{3D}[e,\omega] = \int d^3x \;  \epsilon^{\mu\nu\sigma}\; \big{(} \;\epsilon_{IJKL}\; x^{I}e^{J}_{\mu}F^{KL}_{\nu\sigma} + \frac{1}{\gamma}x^{I}e^{J}_{\mu}F^{KL}_{\nu\sigma}\;\; \big{)} \;\;\;\text{with}\;\;\;  x^{I} = e^{I}_{3}\,.
\end{equation} 
Here the action admits an $SL(2,\mathbb C)$ gauge symmetry.
It is straightforward to show that this action is equivalent to first order three-dimensional gravity. Therefore, $\gamma$ disappears from the classical equations of motion exactly as it does disappear in the Holst action.
As three-dimensional gravity defines an exactly (quantum and classical) solvable system, we can exploit this model to understand deeper the meaning of $\gamma$. This is exactly what has been done in \cite{Achour3,Achour4}.

We proceeded to the Hamiltonian analysis in two different gauges. One choice selects an $SU(1,1)$ non-compact subgroup of $SL(2,\mathbb{C})$, the other one selects a compact subgroup $SU(2)$.
The later choice corresponds to the usual time gauge in the Ashtekar-Barbero formulation of gravity. It turns out that the $SU(1,1)$ phase space is manifestly $\gamma$-independent 
(nor the Poisson bracket nor the Hamiltonian constraint depend on $\gamma$) 
while we recover the usual $\gamma$-dependency for the $SU(2)$ phase space with the non-polynomial Lorentzian part of the Hamitlonian constraint. 
From this result, it seems that the Immirzi parameter keeps trace of the non compactness of the initial gauge group.  

Quantizing  the two-gauged action is very instructive. In the time gauge, the theory has been quantized using the same strategy as in four dimensions. At the kinematical level, quantum geometry states
are $SU(2)$ spin-networks: geometric operators are diagonalized by spin-networks states with a discrete and $\gamma$-dependent spectrum.  As a consequence, everything works as in four dimensions.
The novelty is that the theory is now exactly solvable and one can construct the physical states. This is rather immediate and leads to the fact that $\gamma$ disappears from physical predictions when one solves 
the Hamiltonian constraint. Furthermore, the spectrum of geometric operators becomes continuous. Concerning the quantization of the $SU(1,1)$ gauge-fixed theory, it leads immediately at the kinematical level to
the right $\gamma$-independent continuous spectrum. 

From those results, we see that the presence of $\gamma$ at the kinematical level seems to be a gauge artifact, and the parameter disappears once the hamiltonian constraint is taken into account. It is interesting to note that our analytic continuation prescription $(1.4)$ which was successfully tested in the context of black hole is also the one which maps the two kinematical area spectrums of three dimensional gravity in the two different gauge fixing. In that respect,
everything happens as if $\gamma$ has to be fixed to $\pm i$. At this stage, one could wonder if the same phenomenon occurs in the $3+1$ theory. Only an explicit hamiltonian analysis  of the $3+1$ Holst action in a (yet-to-be-defined) non compact gauge, which would reduce $SL(2,\mathbb{C})$ to $SU(1,1)$, could allow us to answer the question. Although this work was done in the context of three dimensional gravity, it reveals a deep link between the presence of the Immirzi parameter in the kinematical predictions of Loop Quantum gravity and the compactness of the group we are working with.

Finally, this analytic continuation prescription has been recently implemented in Loop Quantum Cosmology for a flat universe with vanishing cosmological constant \cite{Achour5}. It turns out that after applying our prescription, the curvature of the space-time and the energy density remain both bounded which is a non trivial result. It seems therefore that our prescription preserves also the bouncing scenario of Loop Quantum Cosmology. For more details on this new model, see \cite{Achour5}. 

 This analytic continuation constitutes a proposal for defining a theory of self-dual quantum gravity in terms of the complex Ashtekar connection and for solving the so-called reality conditions. We expect that the systematic investigation of this analytic continuation in various setups will eventually lead to new insights on the status of the quantum states of complex Ashtekar gravity. The first attempt to define a Wick rotation in the context of Ashtekar gravity was proposed  in \cite{Thiemann}. Establishing a clear link between the two approaches would inevitably shed some light on the one described in this contribution.

\end{document}